\documentclass[aip,jcp,graphicx,amsmath,amssymb,preprint]{revtex4-1}
\usepackage{graphicx}
\usepackage{array}
\usepackage{bm}
\usepackage{natbib}
\usepackage{rotating}
\usepackage{dcolumn}
\usepackage{outlines}
\usepackage{latexsym}
\usepackage{amsfonts}
\usepackage{amssymb}
\usepackage[usenames,dvipsnames,table]{xcolor}
\usepackage{longtable}
\usepackage{float}
\usepackage{amsmath}
\newcommand{\wav}{cm$^{-1}${} }

\newcommand{\hl}[1]{\textcolor{red}{#1}}
\newcommand{\jg}[1]{\textcolor{green}{#1}}
\newcommand{\kc}[1]{\textcolor{blue}{#1}}
\usepackage[version=3]{mhchem} 

\relax 
\gdef \LT@i {\LT@entry 
    {1}{81.04553pt}\LT@entry 
    {1}{42.93335pt}\LT@entry 
    {3}{49.01671pt}\LT@entry 
    {3}{49.01671pt}\LT@entry 
    {3}{49.01671pt}}
\bibdata{CCl++C2H2Notes,CCl++C2H2}
\bibstyle{aipnum4-1}
\begin{document}

\title{Translationally cold trapped CCl$^+$ reactions with acetylene (C$_2$H$_2$) }

\author{K. J. Catani}
\email{katherine.catani@colorado.edu}
\author{J. Greenberg}
\author{B. V. Saarel}
\author{H. J. Lewandowski}
\affiliation{JILA and the Department of Physics, University of Colorado, Boulder, Colorado, 80309-0440, USA}

\date{\today}

\begin{abstract}
Ion-neutral chemical reactions are important in several areas of chemistry, including in some regions of the interstellar medium, planetary atmospheres, and comets. Reactions of CCl$^+$ with C$_2$H$_2$ are measured and the main products include \emph{c}-C$_3$H$_2$$^+$ and \emph{l}-C$_3$H$^+$, both relevant in extraterrestrial environments. Accurate branching ratios are obtained, which favor formation of \emph{c}-C$_3$H$_2$$^+$ over \emph{l}-C$_3$H$^+$ by a factor of four. Measured rate constants are on the order of Langevin and complementary electronic structure calculations are used to aid in the interpretation of experimental results.

\end{abstract}


\maketitle 

\section{Introduction} \label{intro}

Chlorine containing compounds have been found to exist in extraterrestrial environments including the dense and diffuse clouds of the interstellar medium (ISM). Interest in chlorine chemistry in these areas was initially fueled by detection of HCl, HCl$^+$, and H$_2$Cl$^+$ in these areas. However, these small molecules are acknowledged to account for only a minor fraction of the overall chlorine abundance.\cite{Smith:1985wu,Blake:1986ufa,Zmuidzinas:1995vn,Neufeld:2009,Lis:2010ug,Neufeld:2012hi,Goto:2013cm,Muller:2014je,Lanza:2014dv,Neufeld:2015em} More recently, CH$_3$Cl was detected towards the low mass protostar IRAS 16293-2422 and comet Churyumov-Gerasimento.\cite{Fayolle2017} Organohalogens, including CF$^+$ have been detected in these remote environments, but their overall chemical roles are not completely understood. A possible reservoir containing both carbon and chlorine is the carbon monochloride cation (CCl$^+$). CCl$^+$ has been proposed as an intermediate in the overall chlorine cycle in diffuse clouds, although its overall abundance is not known. The major predicted pathways to CCl$^+$ formation are through reactions of C$^+$ with HCl and H$_2$Cl.\cite{Anicich:1976,Smith:1985wu,Blake:1986ufa,Sonnenfroh:1988tg,Dateo:1989il,Clary:1990vm,Glosik:1993hn,Glosik:1993vu} Previous kinetic measurements, mainly using ion cyclotron resonance (ICR) spectrometry and selected ion flow tube (SIFT) techniques, found CCl$^+$ to be stable and non-reactive.\cite{Smith:1985wu,Blake:1986ufa,Anicich:1993} Products were observed for reactions with NH$_3$, H$_2$CO, CH$_3$Cl, CHCl$_3$, CCl$_4$, CHCl$_2$F, and CHClF$_2$.\cite{Smith:1985wu,Blake:1986ufa,Anicich:1993} However, with NO, H$_2$, CH$_4$, N$_2$, O$_2$, CO, and CO$_2$, no reactions were observed.\cite{Smith:1985wu,Blake:1986ufa,Anicich:1993} Although only a few reactions with relevant interstellar species have been measured, it is conceivable that CCl$^+$ could undergo reactive collisions with varying neutrals, such as acetylene (C$_2$H$_2$; previously detected and thought to be abundant).\cite{Lacy:1989,Lahuis:2000,Mumma:2003,Waite:2007,Agundez:2008}  Reactions of CCl$^+$ with neutral carbonaceous species would likely produce larger carbocation species.

\begin{figure}[htb]
\includegraphics[width=5.5cm]{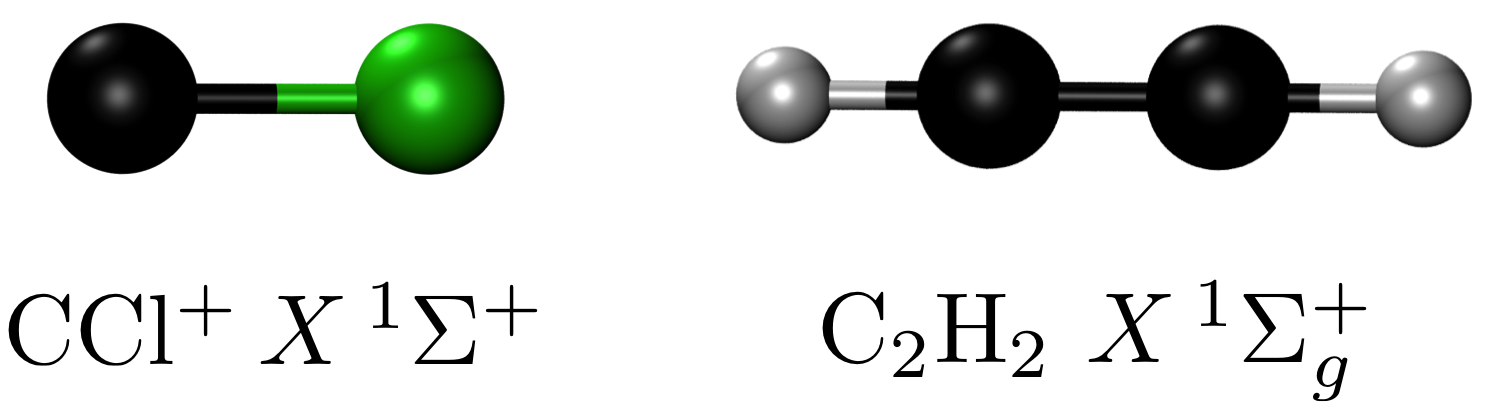}
\caption{Ground-state structures of CCl$^+$ and C$_2$H$_2$.}
\label{ccl}
\end{figure}

Simple carbocation species like C$_2$H$_2$, $c$-C$_3$H$_2$$^+$, and $l$-C$_3$H$^+$ (observed cationic reaction products), have long been speculated to be present in the ISM. Recently, \emph{l}-C$_3$H$^+$ was detected towards dark nebula in Orion and confirmed with laboratory spectroscopic measurements.\cite{Pety:2012,Brunken2014,Botschwina:2014gq,Guzman:2015iv,McGuire:2018hw,Gerin:2019ej} The role of C$_3$H$_2$$^+$ in these environments is less certain. Both neutral \emph{l}-C$_3$H$_2$ and \emph{c}-C$_3$H$_2$ have been detected towards dark clouds in the interstellar medium by their radioemission lines.\cite{Thaddeus:1985} $c$-C$_3$H$_2$, in particular, has been detected in diffuse clouds, where ionizing radiation is more prevalent, and the role of its cation has been speculated as its probable precursor. Insight into additional formation pathways of these fundamental carbocations could prove useful for refining chemical models that rely on energetic and electronic information.

Understanding the details of these and other ion-neutral chemical reactions benefit from studies at low temperatures and are beginning to be explored using new experimental techniques. In particular, trapped Coulomb crystals have become increasingly popular as a way of studying cold ion-neutral chemical reactions.\cite{Chang:2013,Chen:2019,Greenberg:2018,Kilaj:2018,Okada:2013,Petralia:2020,Puri:2017,Rosch:2014,Schmid:2019,Yang:2018}  Here, the reactant ion is sympathetically cooled to low translational temperatures by elastic collisions with co-trapped laser-cooled calcium ions. This technique allows for controlled interrogation of cold ion-neutral reactions over long timescales.

To characterize the CCl$^+$ cation and its reactivity in remote environments, we have measured the reactive pathways of gas-phase CCl$^+$ with C$_2$H$_2$ in a linear Paul ion trap (LIT) coupled to a time-of-flight mass spectrometer (TOF-MS). The coupled ion trap enables co-trapping of sympathetically cooled CCl$^+$ with laser cooled Ca$^+$. This creates a controlled environment to study reactions with C$_2$H$_2$ that may be present is some areas of the ISM, planetary atmospheres, or comets. The reaction of CCl$^+$ + C$_2$H$_2$ leads to a possible pathway to the formation of \emph{c}-C$_3$H$_2$$^+$ and \emph{l}-C$_3$H$^+$ and to a measurement of the branching ratio of these two products. Complementary electronic structure calculations are used to aid in the discussion of the experimental results. 

\section{Experimental Methods}

Kinetic data for the reactions of CCl$^+$ + C$_2$H$_2$ were recorded at different interaction time steps by monitoring the masses of all ions present in a custom built LIT radially coupled to a TOF-MS. Details of the apparatus have been described previously,\cite{Schmid:2017,Greenberg:2018,Schmid:2019} and only a brief outline is presented here. CCl$^+$ was formed through non-resonant multiphoton ionization of tetrachloroethylene (TCE; C$_2$Cl$_4$). Vapor from TCE was seeded in a pulsed supersonic expansion of helium gas ($3\,\%$\,TCE in $\sim$1000\,Torr He). The skimmed molecular beam was overlapped with 216\,nm photons from a frequency-doubled, 10\,Hz pulsed dye laser (LIOP-TEC LiopStar; 10\,ns pulse, 100\,$\mu$J/pulse) in the center of the ion trap forming C$^{35}$Cl$^+$, C$^{37}$Cl$^+$, C$_2$$^+$, C$_2$H$^+$, $^{35}$Cl$^+$, and $^{37}$Cl$^+$. Radio frequency (RF) secular excitation of contaminant ion masses was used to eject all cation species from the trap except C$^{35}$Cl$^+$ cations (subsequently denoted as CCl$^+$). Next, calcium was added to the chamber via effusive beam from a resistively heated oven and non-resonantly photoionized with 355\,nm light from the third harmonic of a Nd:YAG laser (Minilite, 10\,Hz, 7\,mJ/pulse). Trapped calcium ions were Doppler laser cooled with two external cavity diode lasers (ECDLs) forming a Coulomb crystal composed of Ca$^+$ and a pure sample of CCl$^+$. Both ECDLs were fiber-coupled with one tuned to the $^2\text{S}_{1/2}-^2\text{P}_{1/2}$~ transition of Ca$^+$ (397\,nm; New Focus, 3.5\,mW, 600\,$\mu$m beam diameter) and the other to re-pump  Ca$^+$ ions back into the cooling cycle on the $^2\text{P}_{1/2}-^2\text{D}_{3/2}$~ transition (866\,nm; New Focus, 9\,mW, 600\,$\mu$m beam diameter). Both laser frequencies were measured and locked using a wavemeter (HighFinesse/\AA ngstrom WSU-30). 

Upon formation of a mixed-species Coulomb crystal, the RF trapping amplitude was dropped to allow any ions in metastable higher angular-momentum orbits to escape from the trap. This was repeated three times, leaving only cold Ca$^+$ and sympathetically cooled CCl$^+$. During the course of each experimental run, trapped Ca$^+$ ions were monitored continuously via their fluorescence using an intensified CCD camera. CCl$^+$ and other product ions do not fluoresce and were not directly seen in the images, but rather were inferred due to changes in the shape of the Coulomb crystal. Non-fluorescing ions were detected and quantified by ejecting the trapped species into a TOF-MS, which enabled unambiguous determination of chemical formula based on the ion mass-to-charge ratio ($m/z$).  

In a typical experimental run, $\sim100$ CCl$^+$ ions were loaded and sympathetically cooled by $\sim800$ calcium ions. Once the mixed Coulomb crystal was formed, neutral C$_2$H$_2$ ($\sim4.7\%$ in Ar at 298\,K) was leaked into the vacuum system containing the trap at a constant pressure of $\sim4.6(3)\times10^{-9}$\,Torr via a pulsed leak valve scheme.\cite{Jiao:1996} The gas pulse duration was one of multiple neutral gas exposure time steps (0, 5, 10, 20, 40, 80, 160, and 260\,s). At the end of each time step, a mass spectrum was measured and ion numbers for each mass channel were recorded. An experimental run for every time step was repeated four times and the number of ions from each mass channel of interest was averaged at each time step. Each mass channel of interest was normalized by the initial number of CCl$^+$ ions, and the average normalized ion numbers and standard error of the mean are plotted as a function of the different time steps as shown in Figure \ref{rxncurve}.
 
\section{Computational Methods} 
Structures, vibrational frequencies, and relaxed potential energy scans for CCl$^+$ + C$_2$H$_2$ were determined using the M06-2X/aug-cc-pVTZ level of theory. Potential energy surfaces were scanned along bond lengths, rotations, and angles to identify minima and saddle points. The M06-2X stationary points were used for CCSD/aug-cc-pVTZ calculations of structures and harmonic vibrational frequencies. The optimized structures were then used for single point energy calculations at the CCSD(T) level of theory with extrapolation to the complete basis set (CBS) limit using the aug-cc-pVXZ ($\text{X}=\text{T,Q}$) basis sets.\cite{augdunning,ccpvtz,Halkier:1999} The energies were corrected for vibrational zero-point energy from the calculated CCSD/aug-cc-pVTZ level of theory. Energies extrapolated to the CBS limit should account for possible basis set superposition errors. Structures for CCl$^+$ and C$_2$H$_2$ are shown in Figure \ref{ccl}, with further geometric parameters and calculated harmonic vibrational frequencies (CCSD/aug-cc-pVTZ) provided in the Supplemental Information. 

Other computational studies on CCl$^+$ itself recommend using multireference methods.\cite{Nishimura:1983,Peterson:1987,Peterson:1990,Sun:2013} The computational cost of using multireference methods for the larger potential energy surface would be prohibitive, therefore we use the CCSD/aug-cc-pVTZ level of theory, which gives reasonable agreement with previously published computational and experimental data for CCl$^+$ and C$_2$H$_2$ separately.\cite{Nishimura:1983,Gruebele:1986,Peterson:1987,Peterson:1990,Sun:2013} Reported CCSD(T)/CBS energies should be accurate to within 0.04\,eV. Computations were compared to previous theoretical work and experimental data where available for the reactants and products in order to determine the best level of theory to determine accurate energetics. 
DFT and CCSD calculations were performed using both the Gaussian 16 and the Psi4 version 1.3.2 computational packages.\cite{g16,Psi4} Relaxed potential energy scans for the complexes at the M06-2X/aug-cc-pVTZ level of theory were undertaken using Gaussian 16.

\section{Results and Discussion}
\subsection{CCl$^+$ + C$_2$H$_2$ reaction measurements}
Using room temperature (298\,K) C$_2$H$_2$, the average collision energy with translationally cold ($\sim10$\,K) CCl$^+$ is 17\,meV ($\sim$\,197\,K). The excess collision energy should not be a significant source of energy to the reactions as the predicted products, \emph{c}-C$_3$H$_2$$^+$ + Cl and \emph{l}-C$_3$H$^+$ + HCl are -0.15 and -0.07\,eV exothermic, respectively  (Equations \ref{c3h2+} and \ref{c3h+}) based on CCSD(T)/CBS level calculations:

\begin{equation}\label{c3h2+}
\begin{split}
\text{CCl}^+ + \text{C}_2\text{H}_2\,\longrightarrow\,\text{\emph{c}-}\text{C}_3\text{H}_2^+ + \text{Cl} \\
\Delta E = -0.15\,\text{eV}\\
\end{split}
\end{equation}
\begin{equation}\label{lc3h2+}
\begin{split}
\text{CCl}^+ + \text{C}_2\text{H}_2\,\longrightarrow\,\text{\emph{l}-C}_3\text{H}_2^+ + \text{Cl} \\
\Delta E = 0.08\,\text{eV}
\end{split}
\end{equation}
\begin{equation}\label{c3h+}
\begin{split}
\text{CCl}^+ + \text{C}_2\text{H}_2\,\longrightarrow\,\text{\emph{l}-C}_3\text{H}^+ + \text{HCl} \\ 
\Delta E = -0.07\,\text{eV}
\end{split}
\end{equation}

Several other product channels were explored theoretically, however, their endothermicities were significantly greater than the calculated average collision energy (greater than 1\,eV above the reactant energies), and therefore disqualified from the discussion of products. In addition, the \emph{l}-C$_3$H$_2$$^+$ isomer could contribute to the $m/z$ 38 signal, however, the pathway is endothermic by about 0.08\,eV, making it very unlikely to be formed (see Equation \ref{lc3h2+}). 

\begin{figure}[htb]
\includegraphics[width=8.5cm]{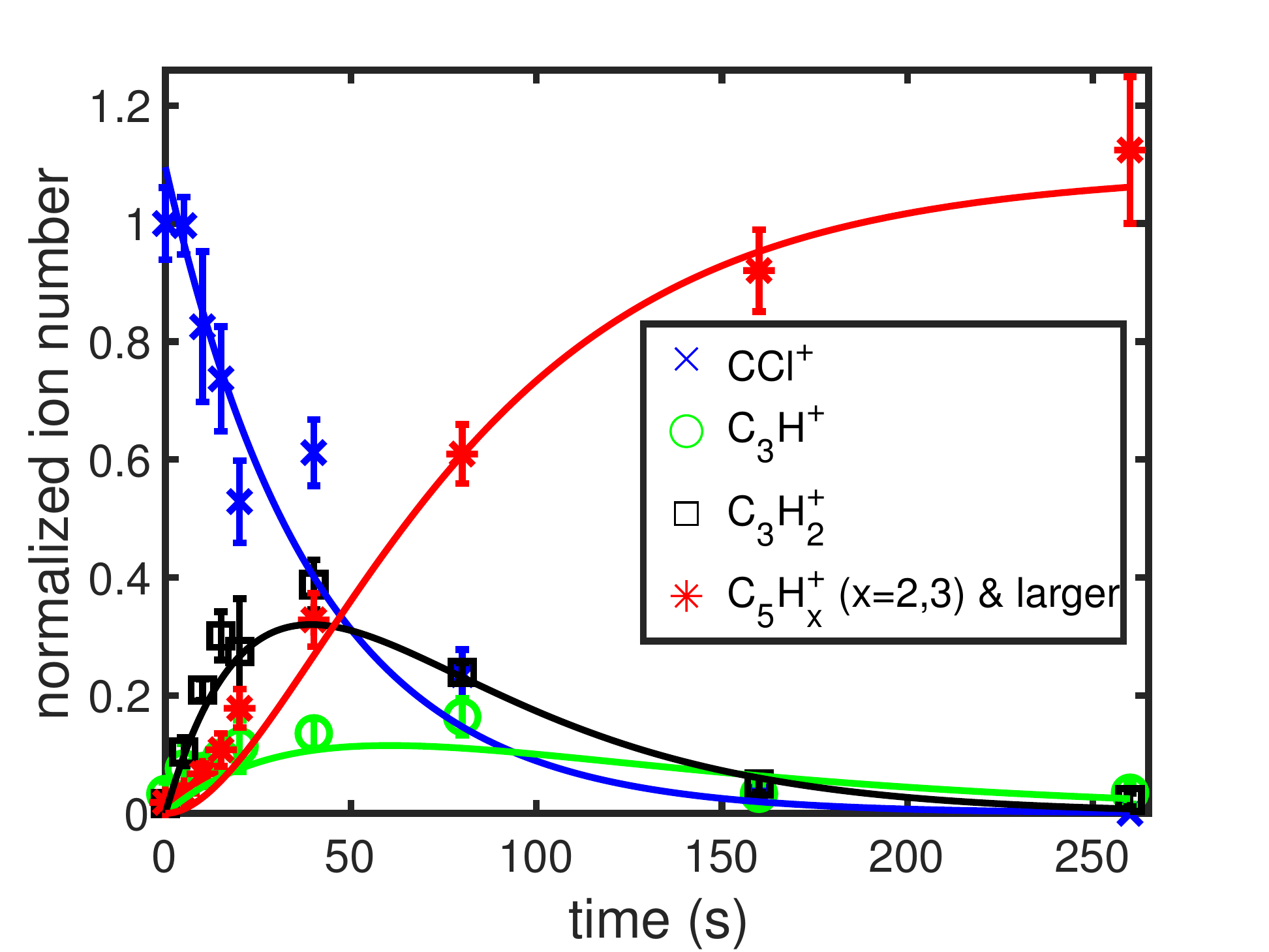}
\caption{Measured ion numbers of CCl$^+$ (blue\,\kc{$\bm{\times}$}), $l$-C$_3$H$^+$ (green\,\jg{$\bm{\circ}$}), $c$-C$_3$H$_2$$^+$ (black\,$\bm{\Box}$), and C$_5$H$_\text{x}$$^+$\,$(\text{x}=2,3$, and higher; red\,\hl{$\bm{\ast}$}) as a function of time. Data are normalized by the initial ion number of CCl$^+$ ($\sim 100$). Each data point represents the mean and standard error from four experimental runs per time point. The averaged data are fit using a pseudo-first order reaction rate model (solid lines).}
\label{rxncurve}
\end{figure}

As shown in Figure \ref{rxncurve}, the initial ionic products that are formed as CCl$^+$ reacts away are indeed C$_3$H$^+$ and C$_3$H$_2$$^+$ ($m/z$ 37 and 38, respectively). At about 40\,seconds, the rates into the intermediate C$_3$H$^+$ and C$_3$H$_2$$^+$ species are overtaken by the rates of these intermediate species reacting with the excess C$_2$H$_2$ present. This is illustrated by the turning over of the two corresponding curves (green and black traces) and growth of the secondary products of larger carbonaceous species (red trace), mainly comprising C$_5$H$_2$$^+$ and C$_5$H$_3$$^+$. The red trace plotted in Figure \ref{rxncurve} includes all mass channels above $m/z$ 61, which should account for any secondary and tertiary reaction products (or larger). Additions of C$_2$H$_2$ to C$_3$H$_\text{x}$$^+$\,$(\text{x}=1,2)$ have been characterized previously, and are not modeled computationally.\cite{Szabo:1971,Schiff:1979,Anicich:1986,Smith:1987,McElvany:1988}

Experimental reaction rates are determined by fitting the reaction data, shown in Figure \ref{rxncurve} to a pseudo-first-order reaction rate model; the fits are represented by solid lines. For the formation of intermediate and secondary products, the model is explicitly defined as: 
\begin{equation}
\frac{d[\text{CCl}^+]}{dt}=-(k_{37}+k_{38})[\text{CCl}^+]
\end{equation}
\begin{equation}
\frac{d[\text{C}_3\text{H}^+]}{dt}=k_{37}[\text{CCl}^+]-k_{37\_63}[\text{C}_3\text{H}^+]
\end{equation}
\begin{equation}
\frac{d[\text{C}_3\text{H}_2^+]}{dt}=k_{38}[\text{CCl}^+]-k_{38\_63}[\text{C}_3\text{H}_2^+]
\end{equation}
\begin{equation}
\frac{d[\text{C}_5\text{H}_\text{x}^+]}{dt}=k_{37\_63}[\text{C}_3\text{H}^+]+k_{38\_63}[\text{C}_3\text{H}_2^+]
\end{equation}
where the product ions of C$_3$H$^+$ (\emph{m/z} 37), C$_3$H$_2$$^+$ (\emph{m/z} 38), and C$_5$H$_\text{x}$$^+$ (x=2,3, and higher; \emph{m/z} $\geq$ 62) at time \emph{t} are normalized with respect to the number of CCl$^+$ ($\sim100$) ions at $t=0$. Rate constants are estimated using the measured concentration of neutral C$_2$H$_2$, which is measured using a Bayard-Alpert ion gauge $\approx5(1)\times10^6$\,cm$^3$. It should be noted that the uncertainty quoted here is statistical, there is larger systematic uncertainty introduced by using an ion gauge in the $10^{-10}$ Torr regime. The resulting reaction rates and rate constants are given in Table \ref{rates}. These calculated rate constants are on the order of the rate calculated using Langevin collision theory, $1.07\times10^{-9}$ cm$^3$s$^{-1}$. 

\begin{table}[htb]
\protect\caption{Results from the fits to the pseudo-first order reaction rate models for the formation of intermediate and secondary products from the reaction of CCl$^+$ + C$_2$H$_2$. Included are the rate constants for the single pressure measurement of C$_2$H$_2$\,$\approx5(1)\times10^6$\,cm$^3$.}\label{rates}
\centering
\begin{tabular}{lccc}
\hline			
\hline			
Species&	$m/z$ & $k$ (s$^{-1}$) 	&	$k/[\text{C}_2\text{H}_2]$ (cm$^3$s$^{-1}$) 	\\
\hline					
C$_3$H$^+$ & 37	&	$0.5(3)\times10^{-2}$	&	$0.9(5)\times10^{-9}$	\\
C$_3$H$_2$$^+$& 38	&	$2.0(3)\times10^{-2}$	&	$4(1)\times10^{-9}$	\\
C$_5$H$_\text{x}$$^+$$^a$ &62 (and larger)	&	$4(1)\times10^{-2}$	&	$7(2)\times10^{-9}$	\\
\hline			
\hline	
\multicolumn{4}{l}{ \tiny{$^a$formation of C$_5$H$_\text{x}$$^+$ and higher order products from both C$_3$H$_2$$^+$ and C$_3$H$^+$}}\\		
\end{tabular}
\end{table}
From the calculated rate constants, the branching ratio between the two intermediate products is determined to be about 4:1, with the Cl loss product channel (formation of \emph{c}-C$_3$H$_2$$^+$) favored over the HCl loss product channel (formation of \emph{l}-C$_3$H$^+$).

Further confirmation of the intermediate and secondary products was obtained by reaction of CCl$^+$ with deuterated acetylene (C$_2$D$_2$), where the product masses shifted to $m/z$ 38 for C$_3$D$^+$ and $m/z$ 40 for C$_3$D$_2$$^+$ as expected. Additionally, for both reactions with C$_2$H$_2$ and C$_2$D$_2$, the total ion numbers as a function of trap time were also collected to ensure conservation of trapped ions. These data are included in the Supplementary Information. 

\subsection{Modeling the CCl$^+$ + C$_2$H$_2$ reaction}
\begin{figure}[htb]
\includegraphics[width=8.5cm]{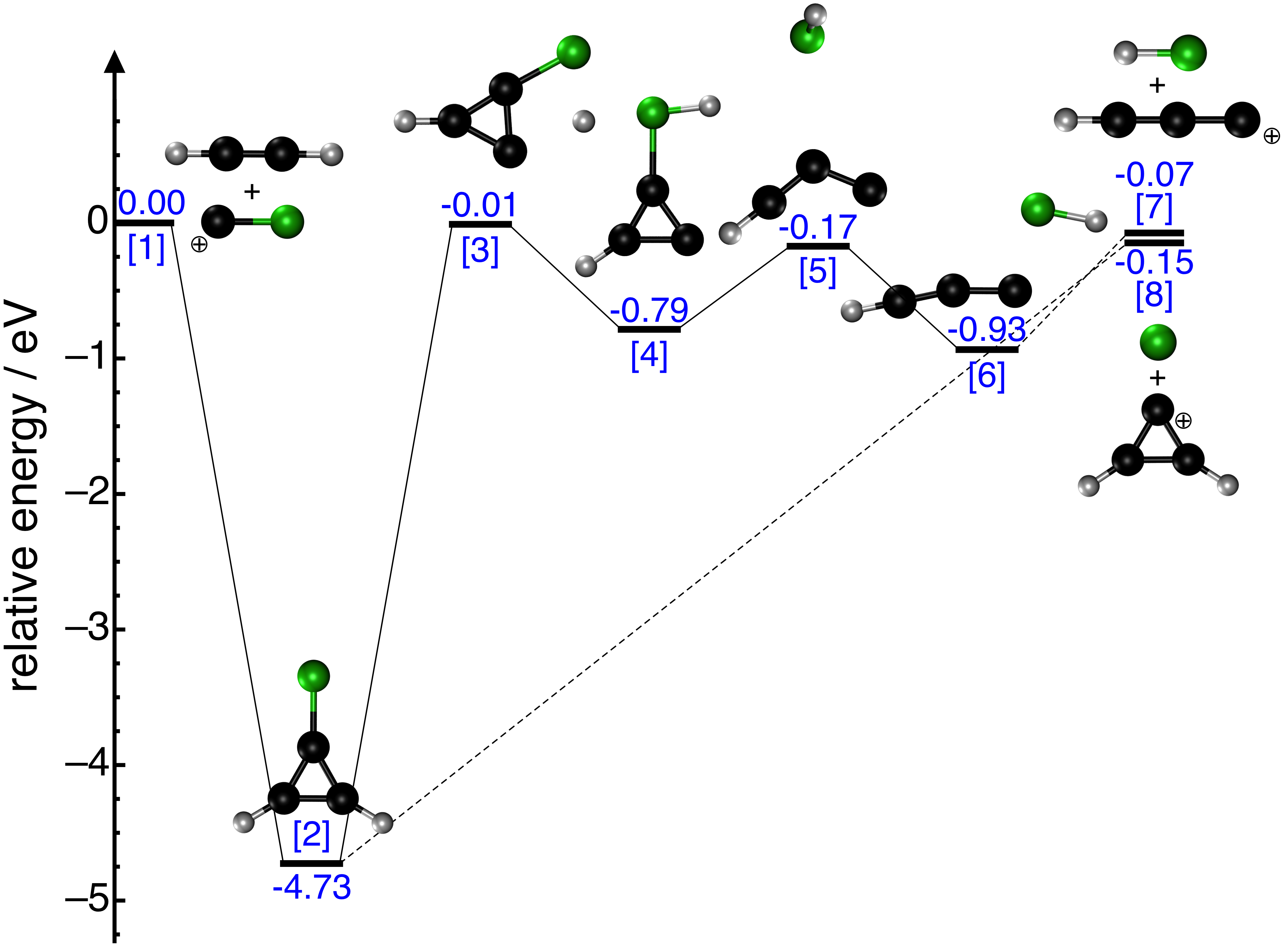}
\caption{The main features of the potential energy surface for CCl$^+$ + C$_2$H$_2$ calculated at the CCSD(T)/CBS level of theory. The favored Cl loss product, dissociates without a barrier, forming $c$-C$_3$H$_2$$^+$ + Cl [8]. The HCl loss channel leading to $l$-C$_3$H$^+$ + HCl [7] involves a few stationary points with lower energies than the dissociation limit of -0.1\,eV. Here, the bare `+' indicates infinite separation.}
\label{PESsurface}
\end{figure}

In order to determine the relative energetics of the different pathways, an abridged potential energy surface of the CCl$^+$ + C$_2$H$_2$ reaction was explored; the key features that lead to the preferable intermediate products are shown in Figure \ref{PESsurface}. Starting from the two reactants, at varying approaches, a stable C$_3$H$_2$Cl$^+$ $C_{2v}$ complex forms, which is exothermic by about 4.73\,eV [CCSD(T)/CBS]. From this complex, the main experimentally observed products correspond to Cl and HCl loss respectively, discussed below. Geometric parameters and energies (CCSD/aug-cc-pVTZ) of each stationary point are given in the Supplementary Information.
\subsubsection{Cl loss}
The favored products, \emph{c}-C$_3$H$_2$$^+$ + Cl, are obtained through a barrierless dissociation pathway, with the products residing about -0.15\,eV below the reactants. The \emph{c}-C$_3$H$_2$$^+$ isomer is favored over the \emph{l}-C$_3$H$_2$$^+$ isomer by about 0.23\,eV (22\,kJ/mol) and the barrier to rearrangement from linear to cyclic has been calculated in previous studies to be between 1.62-1.75\,eV.\cite{Wong:1993} Indeed, it may be possible that both isomers are present, however, as noted above the \emph{l}-C$_3$H$_2$$^+$ + Cl products are slightly endothermic, by about 0.08\,eV, making it very unlikely to be formed. 

\subsubsection{HCl loss} 
The other intermediate product channel, forming \emph{l}-C$_3$H$^+$ through HCl loss, goes through the same cyclic intermediate as the Cl loss channel [2]. From the cyclic minima, several rearrangements are required before eventually losing neutral HCl from \emph{l}-C$_3$H$^+$. A few shallow minima were found where the HCl moiety rearranges along the conjugated pi-system of \emph{l}-C$_3$H$^+$, the lowest energy isomer is shown [6]. The structures and energies (CCSD/aug-cc-pVTZ) of these different isomers are given in the Supplementary Information. 

The experimentally observed branching ratio favors \emph{c}-C$_3$H$_2$$^+$ over \emph{l}-C$_3$H$^+$ by about a factor of four. In the absence of Rice-Ramsperger-Kassel-Marcus (RRKM) theory/master equation simulations, it is difficult to probe the potential energy surface barriers and their possible contribution to this observed branching ratio between the two primary product channels. The comparison is supported by the increased number of steps required for the HCl loss channel compared to the Cl loss channel. It should be noted that the barrier height of the H abstraction before HCl loss ([3] in Figure \ref{PESsurface}) is variable based on the level of theory employed, with the CCSD(T)/CBS level giving the largest barrier. The uncertainty in the level of theory (0.04\,eV) could push this barrier slightly endothermic, which would kinetically favor the Cl loss channel.

\section{Conclusions}
The reaction of C$_2$H$_2$ with sympathetically cooled CCl$^+$ was examined using a linear Paul ion trap coupled to a TOF-MS. Although CCl$^+$ has been previously predicted to be non-reactive, it is clear that collisions with C$_2$H$_2$ result in either Cl or HCl loss. The primary products formed from these two loss channels are the astrochemically relevant \emph{c}-C$_3$H$_2$$^+$ and \emph{l}-C$_3$H$^+$, respectively. The rate constants measured for the formation of the two primary products are on the order of those predicted through Langevin modeling. Furthermore, the Cl loss channel, occurring through a barrierless dissociation pathway, is favored by a factor of four compared to HCl loss channel, which requires several rearrangements before losing HCl.  

The theoretical work presented in this study would benefit from further exploration at a higher level of theory, possibly with multireference methods to characterize CCl$^+$ and its initial complexation with C$_2$H$_2$. In addition, kinetic modeling, possibly using RRKM theory, would be beneficial for a comprehensive comparison to experimentally measured branching ratios.

In future work, we plan to explore further the reactivity of CCl$^+$ with other astrochemically relevant neutrals. In addition, a natural extension to this work would be to study similar reactions of CCl$^+$ with more control over the neutral reactant, using state-selected molecular ions from a traveling wave Stark decelerator.\cite{Shyur:2018,Shyur:2018a} This will allow further study of ion-neutral reactions in low temperature regimes, where the internal quantum states and external motion can be controlled with high precision.

\begin{acknowledgments}
 This work was supported by the National Science Foundation (PHY-1734006, CHE-1900294) and the Air Force Office of Scientific Research  (FA9550-16-1-0117). 
\end{acknowledgments}

\section*{Data Availability}

The data that support the findings of this study are available in the supplementary information and from the corresponding author upon reasonable request.

\section*{Supplemental Information}

\noindent Data included in this supplementary information consist of reaction curves from reactions of CCl$^+$ + acetylene (C$_2$H$_2$) and CCl$^+$ + deuterated acetylene (C$_2$D$_2$; used to verify reaction products). Also included are measured total ion numbers (a sum of all detected ion channels in a given TOF-MS trace) for each reaction, which are used to demonstrate conservation of ions in the linear Paul ion trap over the entire reaction. The full calculated potential energy surface for the discussion of the CCl$^+$ + C$_2$H$_2$ reaction is included at the CCSD/aug-cc-pVTZ level of theory with additional structural and energetic information provided for each stationary point. 

\begin{figure}[H]
\centering
  \includegraphics[width=10cm]{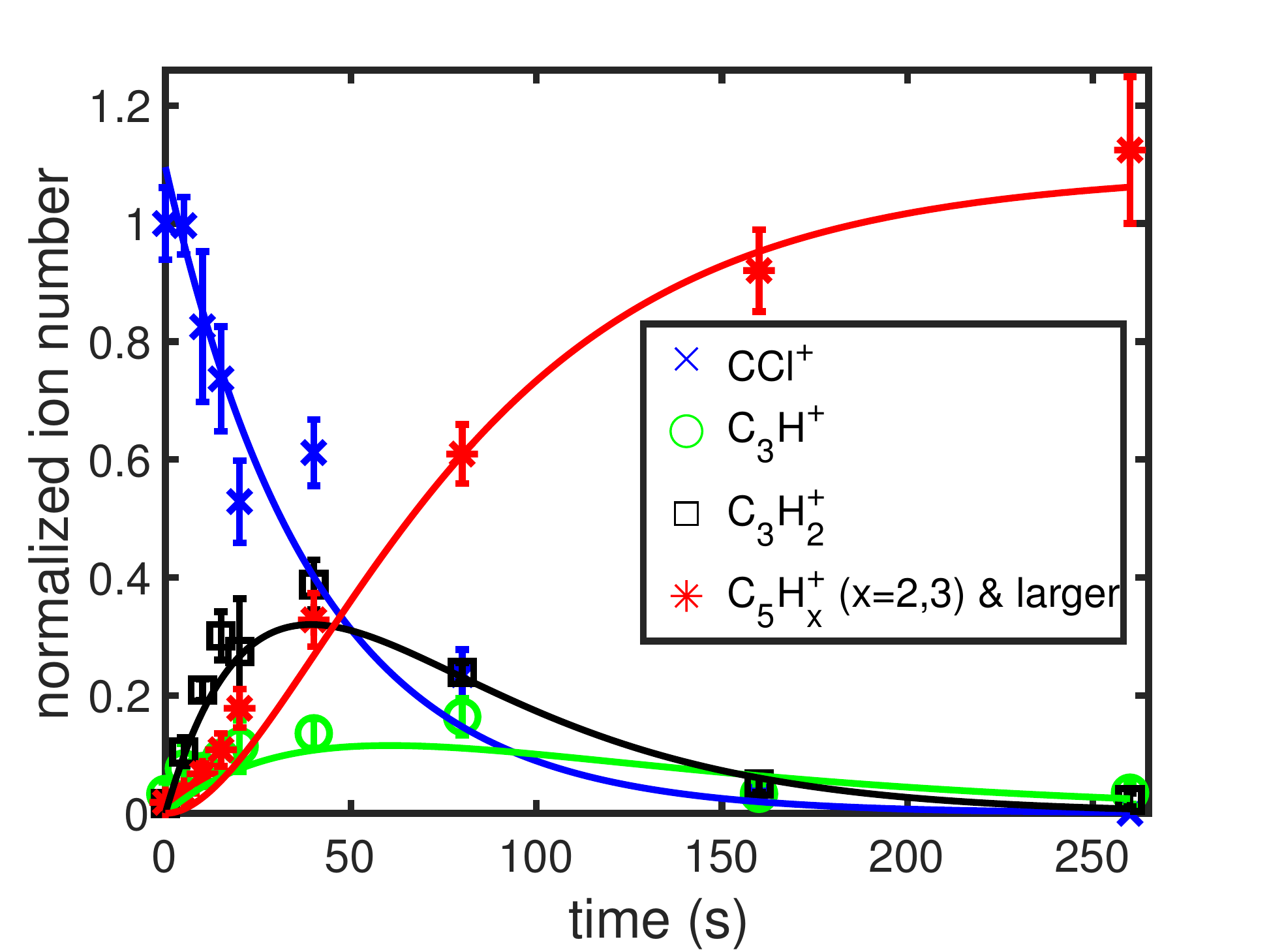}
  \caption{Measured ion numbers of CCl$^+$ (blue $\kc{\bm\times}$), C$_3$H$^+$ (green $\jg{\bm{\circ}}$), C$_3$H$_2$$^+$ (black\,$\bm{\Box}$), and C$_5$H$_\text{x}$\,$(\text{x}=2,3$ and higher; red\,\hl{$\bm{\ast}$}) as a function of time. Data are normalized by the initial ion numbers of CCl$^+$ ($\sim 100$). Each data point represents the mean and standard error from 4 experimental runs per time point. The average data are fit using a pseudo-first order reaction rate model (solid lines). This is Figure \ref{rxncurve} in the main text.}\label{figs1}
\end{figure}

\begin{figure}[H]
\centering
  \includegraphics[width=9.5cm]{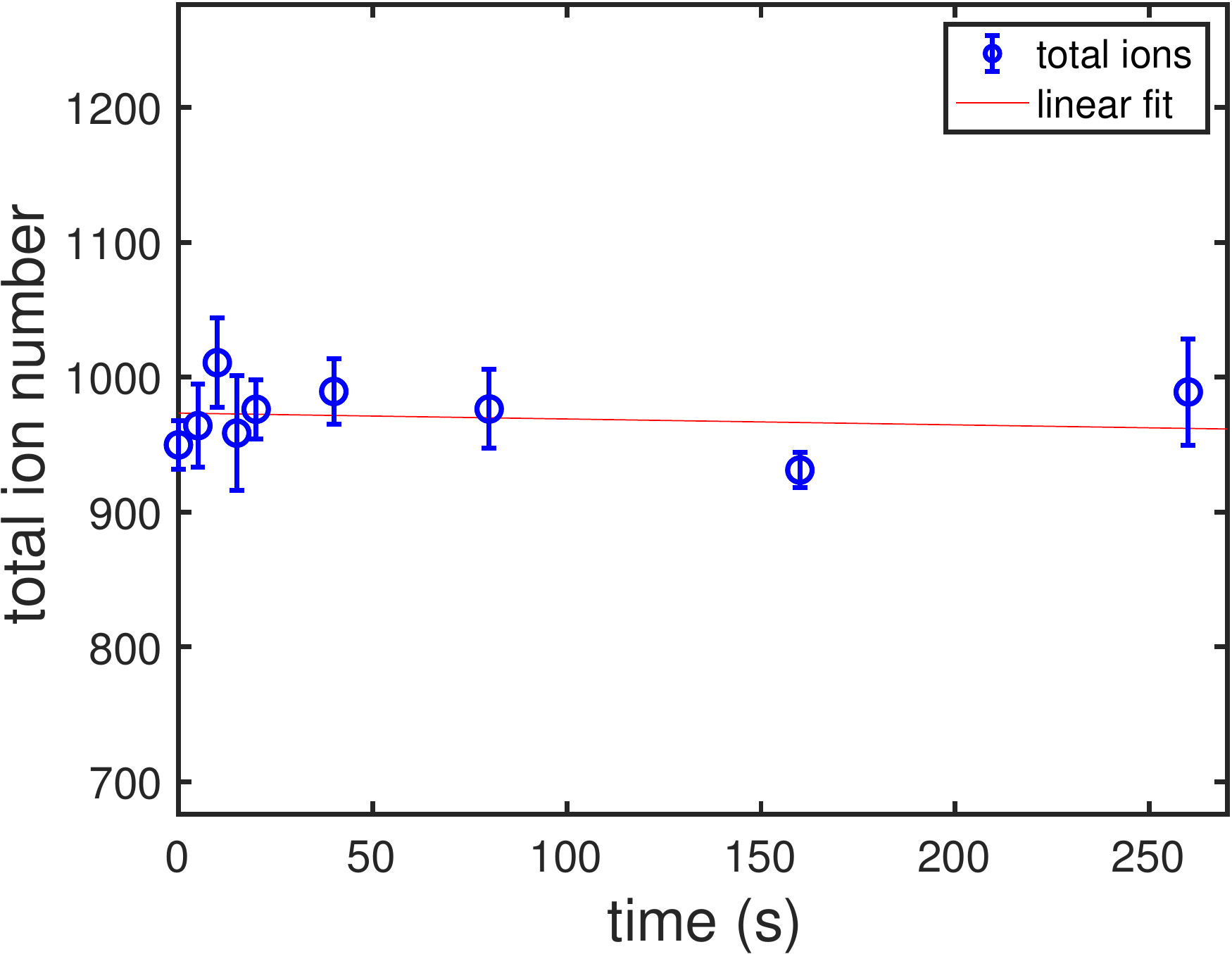}
  \caption{Measured total ion numbers of the CCl$^+$ + C$_2$H$_2$ reaction as a function of time. Total ion numbers represent the sum of all detected ion channels in a given TOF-MS trace and each data point represents the mean and standard error from 4 experimental runs per time point. Total ion data are fit using a linear fit giving a slope of -0.04 with an uncertainty of 0.3 (the 95\% confidence interval of the fit parameter) indicating the number of ions are conserved over the reaction.}\label{figs2}
\end{figure}

\begin{figure}[H]
\centering
  \includegraphics[width=10cm]{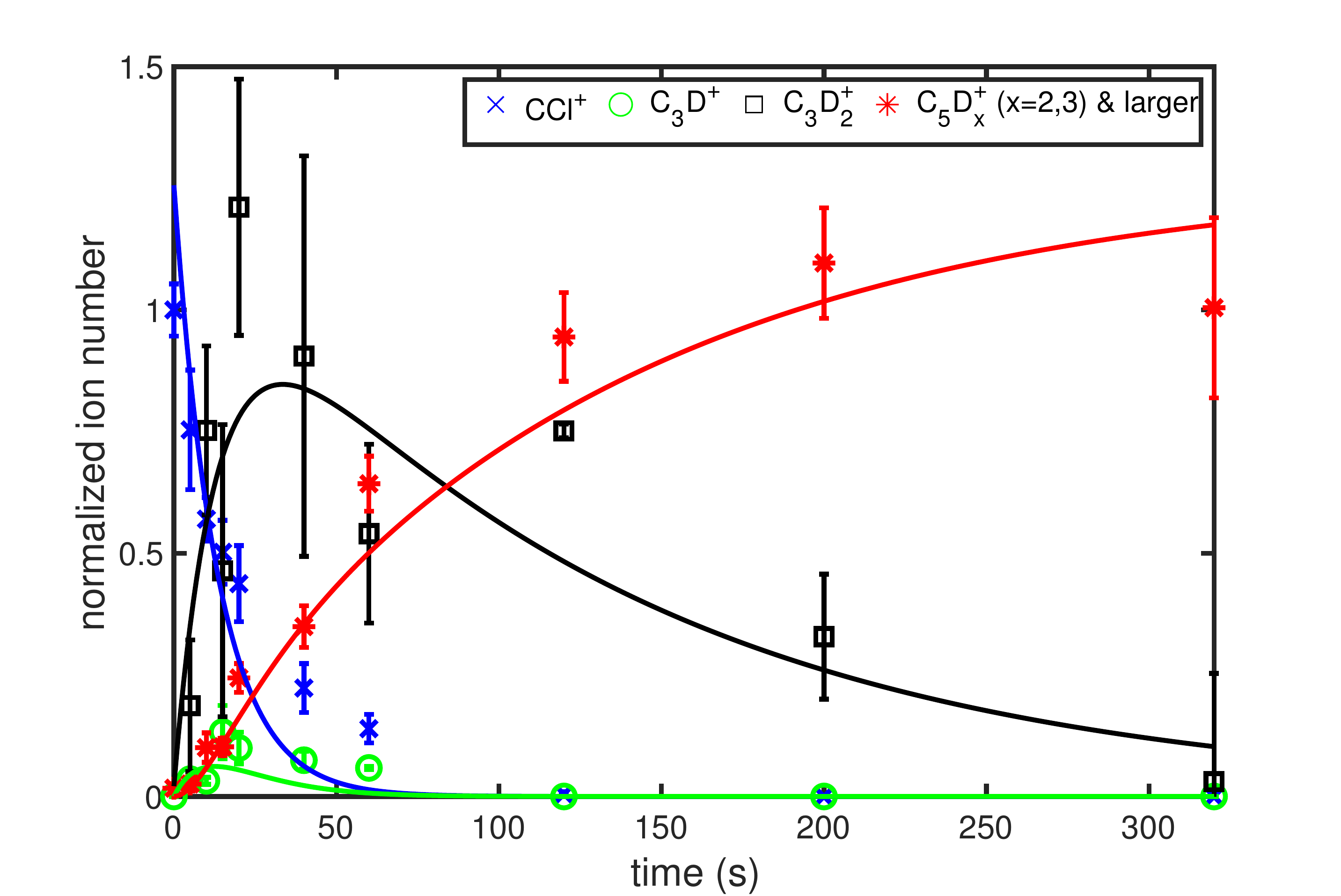}
  \caption{Measured ion numbers of CCl$^+$ (blue $\kc{\bm\times}$), C$_3$D$^+$ (green $\jg{\bm{\circ}}$), C$_3$D$_2$$^+$ (black\,$\bm{\Box}$), and C$_5$D$_\text{x}$\,$(\text{x}=2,3$ and higher; red\,\hl{$\bm{\ast}$}) as a function of time for the reaction of CCl$^+$ with deuterated acetylene (C$_2$D$_2$). Data are normalized by the initial ion numbers of CCl$^+$ ($\sim 100$). Each data point represents the mean and standard error from 4 experimental runs per time point. The average data are fit using a pseudo-first order reaction rate model (solid lines). The reaction traces clearly show that the product masses shift as expected due to deuterium substitution. C$_3$D$_2$$^+$ (\emph{m/z} 40) overlaps with the Ca$^+$ channel (also \emph{m/z} 40). To fit this channel, the initial average Ca$^+$ ion numbers are subtracted from the $m/z$ 40 channel at each time step. There is some variability in the loaded numbers of Ca$^+$, making the ion numbers and rates extracted from the 40 channel less certain.}\label{figs3}
\end{figure}

\begin{figure}[H]
\centering
  \includegraphics[width=9.5cm]{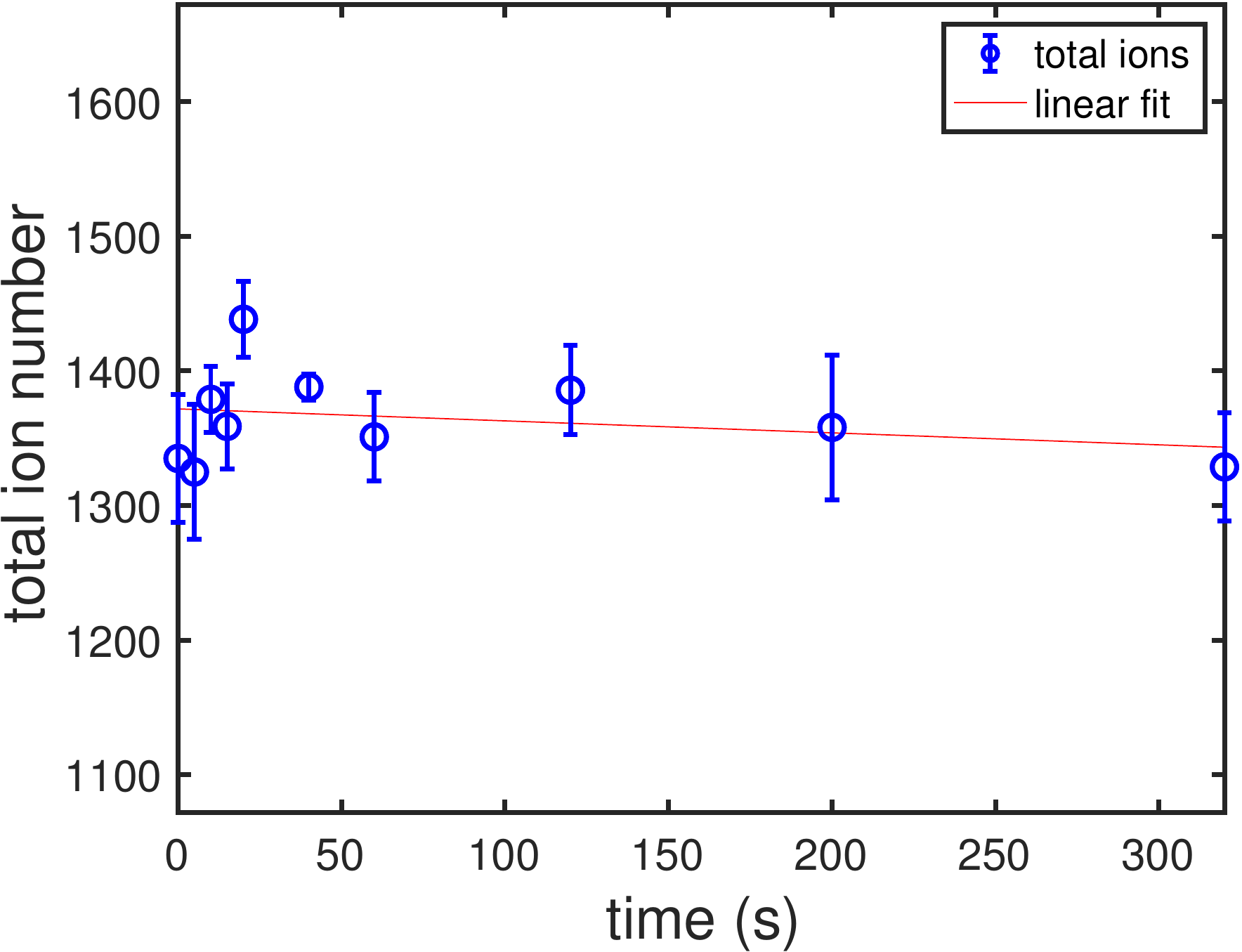}
  \caption{Measured total ion numbers of the CCl$^+$ + C$_2$D$_2$ reaction as a function of time. Each data point represents the mean and standard error from 4 experimental runs per time point. The raw total ion data are fit using a linear fit giving a slope of -0.09 with an uncertainty of 0.2 (the 95\% confidence interval of the fit parameter) indicating the number of ions are conserved over the reaction.}\label{figs4}
\end{figure}

\begin{table}[H]
\caption{Results from the fits to the pseudo-first order reaction rate models for the formation of intermediate and secondary products from the reaction of CCl$^+$ with C$_2$H$_2$ (from Table \ref{rates} in main text) and C$_2$D$_2$. Included are the rate constants for the single pressure measurement of C$_2$H$_2\approx5(1)\times10^6$\,cm$^3$ and C$_2$D$_2\approx7(1)\times10^6$\,cm$^3$. For the C$_2$D$_2$ case, rates extracted from the data are less certain and the majority of the reported error is due to the calcium ion number uncertainties, as the C$_3$D$_2$$^+$ product channel overlaps with the Ca$^+$ channel.}\label{tabs1}
\centering
\begin{tabular}{lccc}
\multicolumn{4}{l}{CCl$^+$ + C$_2$H$_2$} \\
\hline			
\hline			
Species&	$m/z$&$k$ (s$^{-1}$)	&	$k/[\text{C}_2\text{H}_2]$ (cm$^3$s$^{-1}$) 	\\
\hline					
C$_3$H$^+$ & 37	&	$0.5(3)\times10^{-2}$	&	$0.9(5)\times10^{-9}$	\\
C$_3$H$_2$$^+$& 38	&	$2.0(3)\times10^{-2}$	&	$4(1)\times10^{-9}$	\\
C$_5$H$_\text{x}$$^+$$^a$ &62 (and larger)	&	$4(1)\times10^{-2}$	&	$7(2)\times10^{-9}$	\\
\hline
\hline
\multicolumn{4}{l}{CCl$^+$ + C$_2$D$_2$} \\
\hline
\hline
Species&	$m/z$&$k$ (s$^{-1}$) 	&	$k/[\text{C}_2\text{D}_2]$ (cm$^3$s$^{-1}$)	\\
\hline
C$_3$D$^+$ & 38	&	$1(1)\times10^{-2}$	&	$1(2)\times10^{-9}$	\\
C$_3$D$_2$$^+$& 40	&	$6(2)\times10^{-2}$	&	$8(4)\times10^{-9}$	\\
C$_5$D$_\text{x}$+$^b$ &64 (and larger)	&	$9(24)\times10^{-2}$	&	$1(3)\times10^{-8}$	\\
\hline			
\hline	
\multicolumn{4}{l}{ \tiny{$^a$formation of C$_5$H$_\text{x}$$^+$ and higher order products from both C$_3$H$_2$$^+$ and C$_3$H$^+$}}\\	
\multicolumn{4}{l}{ \tiny{$^b$formation of C$_5$D$_\text{x}$$^+$ and higher order products from both C$_3$D$_2$$^+$ and C$_3$D$^+$}}\\		
\end{tabular}
\end{table}

\begin{table}[H]
\protect\caption{Previously measured and calculated constants of CCl$^+$ for comparison with current computational results.}\label{s2}
\centering
\small
\begin{tabular}{lccccc}
\hline								
\hline 							
CCl$^+$	&r$_e$\,(\AA)	&	$\omega_e$\,(\wav) & $D_e$\,(eV)	& Ref\\
\hline								
$\mathbf{X^+\,^1\Sigma^+}$~~...$(5\sigma)^2(6\sigma)^2(2\pi_x)^2(2\pi_y)^2(7\sigma)^2$&&&\\
Exp. 	&	1.5378&	1178	&6.55&Ref. \citenum{Gruebele:1986}\\	
MRCI&1.5421&1172&&Ref. \citenum{Peterson:1990}\\
M06-2X/aug-cc-pVTZ$^a$&1.5341&1212&\\
CCSD/aug-cc-pVTZ$^a$ &1.5330 &1169&6.59&\\
\hline
\hline
\multicolumn{5}{l}{ $^a$\scriptsize{this work}}\\

\\
\end{tabular}
\\
\end{table}

\begin{table}[H]
\protect\caption{Electronic energies and vibrational zero-point corrections from calculations at the CCSD/aug-cc-pVTZ level of theory with accompanying CCSD(T)/CBS single point energies, all in Hartrees. Stationary point numbers correspond to those on the CCl$^+$ + C$_2$H$_2$ potential energy surface in Figure \ref{figs5}. Structures 6 and 10-12 are energetically similar isomers where the HCl moiety rearranges along the \emph{l}-C$_3$H$^+$ backbone. The other linear isomer of C$_3$H$_2$$^+$ is also included for reference. All structures are illustrated in Figure \ref{figs6}.}\label{tabs3}
\centering
\begin{tabular}{lccc}
\hline
\hline
Stationary Point	&	Electronic Energy 	&	Vibrational ZPE correction	&	Single Point Energy	\\
	&	CCSD/aug-cc-pVTZ	&	CCSD/aug-cc-pVTZ	&	CCSD(T)/CBS	\\
\hline							
$[1\text{a}]$ CCl$^+$ 	&	-497.3093456	&	0.0027851	&	-497.4439045	\\
$[1\text{b}]$ C$_2$H$_2$ 	&	-77.2104448	&	0.0271492	&	-77.2986219	\\
$[2]$	&	-574.7086427	&	0.0375602	&	-574.9238578	\\
$[3]$	&	-574.5202927	&	0.0294604	&	-574.7423639	\\
$[4]$	&	-574.5572188	&	0.0326468	&	-574.7740926	\\
$[5]$	&	-574.5254313	&	0.0277394	&	-574.7466993	\\
$[6]$	&	-574.5569850	&	0.0310633	&	-574.7779988	\\
$[7\text{a}]$ HCl	&	-460.3686914	&	0.0068998	&	-460.4594045	\\
$[7\text{b}]$ \emph{l}-C$_3$H$^+$ 	&	-114.1521467	&	0.0201516	&	-114.2829693	\\
$[8\text{a}]$ Cl 	&	-459.7017467	&	0.0000000	&	-459.7883079	\\
$[8\text{b}]$ \emph{c}-C$_3$H$_2$$^+$ 	&	-114.8378178	&	0.0333737	&	-114.9630686	\\
$[9]$	 \emph{l}-C$_3$H$_2$$^+$&	-114.8212372	&	0.0265948	&	-114.9479331	\\
$[10]$	&	-574.5530070	&	0.0306167	&	-574.7741961	\\
$[11]$	&	-574.5339041	&	0.0311430	&	-574.7528099	\\
$[12]$	&	-574.5325391	&	0.0306695	&	-574.7513854	\\
\\
\end{tabular}
\\
\end{table}

\begin{figure}[H]
\centering
  \includegraphics[width=15cm]{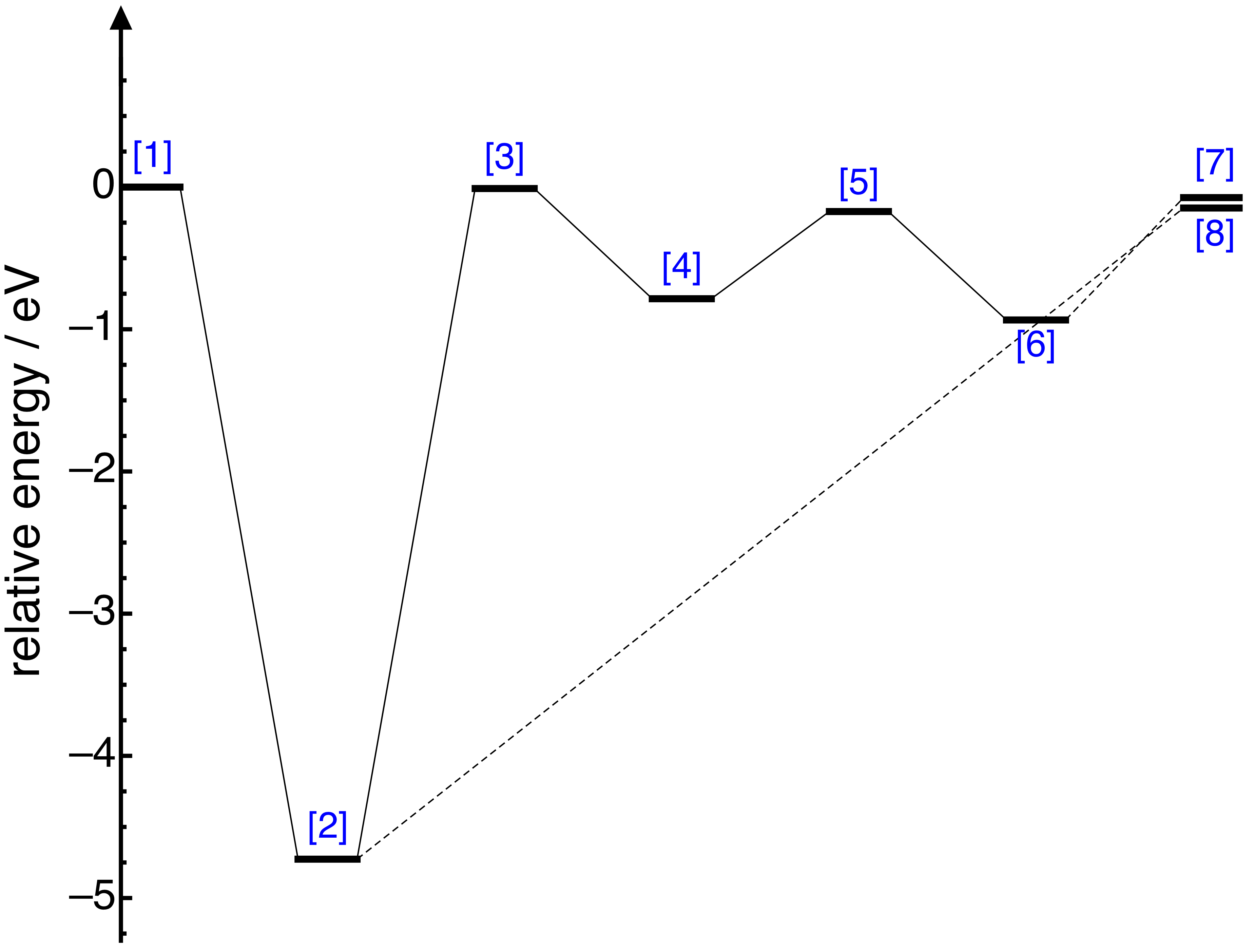}  
  \caption{Potential energy surface calculated at the CCSD/aug-cc-pVTZ level of theory for the reaction of CCl$^+$ with C$_2$H$_2$. Energies (in eV) are from single point CCSD(T)/aug-cc-pVTZ calculations extrapolated to the complete basis set limit. Structures, energies, and geometries are given in Figure \ref{figs6} and Tables \ref{tabs3} and \ref{tabs4}, respectively.}\label{figs5}\end{figure}

\begin{figure}[H]
\centering
  \includegraphics[width=15cm]{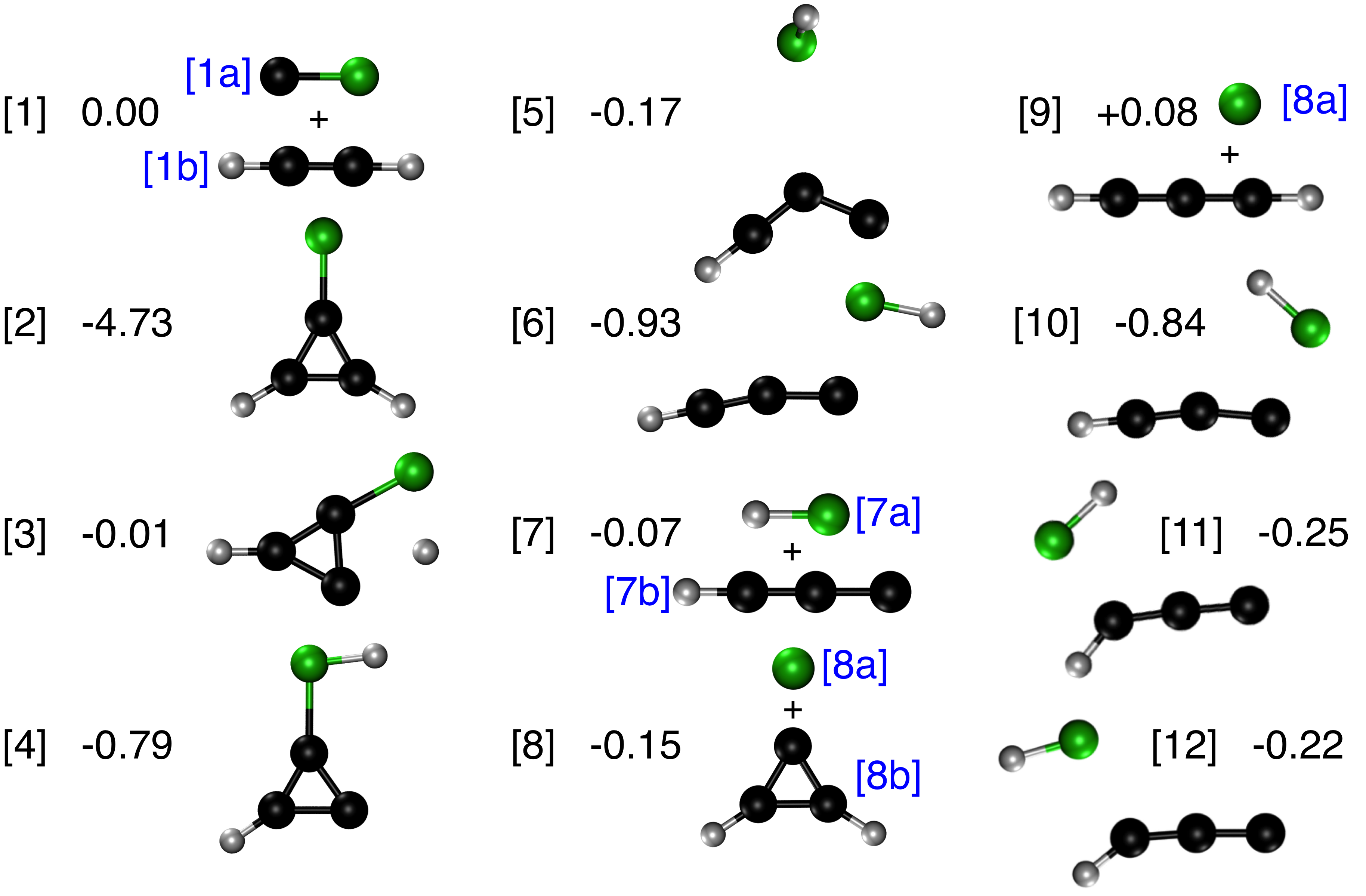}  
  \caption{Structures of stationary points on the potential energy surface in Figure \ref{figs5}. In each case, energies are given relative to the CCl$^+$ + C$_2$H$_2$ limit [1]. Geometries were computed at the CCSD/aug-cc-pVTZ level. Structures 6 and 10-12 are the several minima with different rearrangements of the HCl moiety along the C$_3$H$^+$ backbone. Also included is the \emph{l}-C$_3$H$_2$$^+$ isomer + Cl limit energy. Geometries for each stationary point are provided in Table \ref{tabs3}.} \label{figs6}
\end{figure}

\bigskip

\begin{longtable}[c]{ccccc}
	\caption{CCSD/aug-cc-pVTZ calculated geometries for stationary points on potential energy surface represented in Figure \ref{figs5}. Structures for each point are depicted in Figure \ref{figs6} and electronic and zero point energies are provided in Table \ref{tabs3}.} \label{tabs4} \\
	\small stationary point	&\small ~~~~atom	&	 \small x&	\small y	& \small z	\\ 
	\hline
	\endfirsthead
	\small stationary point	& \small ~~~~atom	&	 \small x&	\small y	& \small z\\
	\hline
	\endhead						
	\small	[1a]	&	\small	C	&	\small	0	&	\small	0	&	\small	-1.14136	\\
	\small		&	\small	Cl	&	\small	0	&	\small	0	&	\small	0.39167	\\
 	\hline													
	\small	[1b]	&	\small	C	&	\small	0	&	\small	0	&	\small	0.599904	\\
	\small		&	\small	H	&	\small	0	&	\small	0	&	\small	1.656109	\\
	\small		&	\small	C	&	\small	0	&	\small	0	&	\small	-0.599904	\\
	\small		&	\small	H	&	\small	0	&	\small	0	&	\small	-1.656109	\\
	\hline													
	\small	[2]	&	\small	C	&	\small	0	&	\small	0	&	\small	-0.345000	\\
	\small		&	\small	Cl	&	\small	0	&	\small	0	&	\small	1.286760	\\
	\small		&	\small	C	&	\small	0	&	\small	0.678367	&	\small	-1.527787	\\
	\small		&	\small	H	&	\small	0	&	\small	1.600247	&	\small	-2.078554	\\
	\small		&	\small	C	&	\small	0	&	\small	-0.678367	&	\small	-1.527787	\\
	\small		&	\small	H	&	\small	0	&	\small	-1.600247	&	\small	-2.078554	\\
	\hline													
	\small	[3]	&	\small	C	&	\small	0.018354	&	\small	0.536955	&	\small	0	\\
	\small		&	\small	Cl	&	\small	-0.768290	&	\small	-0.963326	&	\small	0	\\
	\small		&	\small	C	&	\small	1.394966	&	\small	0.458316	&	\small	0	\\
	\small		&	\small	H	&	\small	0.755356	&	\small	-1.120267	&	\small	0	\\
	\small		&	\small	C	&	\small	0.703112	&	\small	1.675055	&	\small	0	\\
	\small		&	\small	H	&	\small	0.702256	&	\small	2.750014	&	\small	0	\\
	\hline													
	\small	[4]	&	\small	C	&	\small	0.005983	&	\small	0.474440	&	\small	0	\\
	\small		&	\small	Cl	&	\small	-0.085875	&	\small	-1.279367	&	\small	0	\\
	\small		&	\small	C	&	\small	0.860059	&	\small	1.562082	&	\small	0	\\
	\small		&	\small	C	&	\small	-0.594289	&	\small	1.627523	&	\small	0	\\
	\small		&	\small	H	&	\small	-1.455953	&	\small	2.265192	&	\small	0	\\
	\small		&	\small	H	&	\small	1.199848	&	\small	-1.501703	&	\small	0	\\
	\hline													
	\small	[5]	&	\small	C	&	\small	-1.162870	&	\small	0.048628	&	\small	0.038439	\\
	\small		&	\small	Cl	&	\small	1.682021	&	\small	-0.086273	&	\small	-0.050671	\\
	\small		&	\small	C	&	\small	-1.744158	&	\small	1.243183	&	\small	-0.008853	\\
	\small		&	\small	H	&	\small	2.149974	&	\small	0.074659	&	\small	1.128091	\\
	\small		&	\small	C	&	\small	-1.954728	&	\small	-0.900496	&	\small	0.022612	\\
	\small		&	\small	H	&	\small	-2.623554	&	\small	-1.740519	&	\small	0.008556	\\
	\hline
	\small	[6]	&	\small	C	&	\small	1.223681	&	\small	0.310537	&	\small	0	\\
	\small		&	\small	Cl	&	\small	-1.280714	&	\small	-0.425591	&	\small	0	\\
	\small		&	\small	C	&	\small	0.019856	&	\small	0.951579	&	\small	0	\\
	\small		&	\small	H	&	\small	-2.290150	&	\small	0.383234	&	\small	0	\\
	\small		&	\small	C	&	\small	2.394211	&	\small	-0.025809	&	\small	0	\\
	\small		&	\small	H	&	\small	3.413458	&	\small	-0.336866	&	\small	0	\\
	\hline													
	\small	[7a]	&	\small	Cl	&	\small	0	&	\small	0	&	\small	0.035656	\\
	\small		&	\small	H	&	\small	0	&	\small	0	&	\small	-1.237164	\\
	\hline
	\pagebreak													
	\small	[7b]	&	\small	C	&	\small	0	&	\small	0	&	\small	1.197535	\\
	\small		&	\small	H	&	\small	0	&	\small	0	&	\small	2.270605	\\
	\small		&	\small	C	&	\small	0	&	\small	0	&	\small	-0.027009	\\
	\small		&	\small	C	&	\small	0	&	\small	0	&	\small	-1.361224	\\
	\hline													
	\small	[8a]	&	\small	Cl	&	\small	0	&	\small	0	&	\small	0	\\
	\hline
	\small	[8b]	&	\small	C	&	\small	0	&	\small	0.693545	&	\small	0.333591	\\
	\small		&	\small	C	&	\small	0	&	\small	-0.693545	&	\small	0.333591	\\
	\small		&	\small	C	&	\small	0	&	\small	0	&	\small	-0.821531	\\
	\small		&	\small	H	&	\small	0	&	\small	-1.595173	&	\small	0.918901	\\
	\small		&	\small	H	&	\small	0	&	\small	1.595173	&	\small	0.918901	\\
	\end{longtable}

\end{document}